# Correlation between spin-phonon coupling and magneto-electric effects in $CoFe_2O_4$/PMN-PT nanocomposite: Raman Spectroscopy and XMCD study

Anju Ahlawat[1,2*], Azam Ali Khan[1,3], Pratik Deshmukh[1,3], Mandar M. Shirolkar[4], A.K. Sinha[5], S. Satapathy[1,3*], V.G.Sathe[2], R. J. Choudhary[2*]

[1]Laser and Functional Materials Division, Ra ja Ramanna Centre for Advanced Technology, Indore 452013, India
[2]UGC DAE Consortium for Scientific Research, Indore 452001, India
[3]Homi Bhabha National Institute, Training School Complex, Anushakti Nagar, Mumbai-400094, India
[4]Symbiosis Center for Nanoscience and Nanotechnology (SCNN), Symbiosis International (Deemed University) (SIU), Lavale, Pune 412115, Maharashtra, India
[5]Synchrotrons Utilization Section, Raja Ramanna Centre for Advanced Technology, Indore 452017, India
*Corresponding author:anjahlawat@gmail.com*

## Abstract

We have investigated the coupling of lattice with spin via strain interactions in the $CoFe_2O_4$/0.65Pb $(Mg_{1/3}Nb_{2/3})O_3$–$0.35PbTiO_3$(PMN-PT) composite system. X-ray diffraction and Raman spectroscopic studies illustrate a remarkable modification in $CoFe_2O_4$ lattice across Curie temperature ($T_c \sim 450$ K) of PMN-PT. Subsequently, $CoFe_2O_4$/PMN-PT composite reveals a sudden drop in magnetic moment across $T_c$ of PMN-PT (~450 K). However, the independent $CoFe_2O_4$ phase displays typical ferromagnetic behaviour across this temperature. These findings establish spin-lattice coupling owing to the interfacial starin transfer between $CoFe_2O_4$ and PMN-PT in composite. The strain intractions leads to magneto-electric coupling, evidenced by measuring magetization and magneto-electric coefficient for the electric field poled and unploed $CoFe_2O_4$/PMN-PT composite samples. X-ray magnetic circular dichroism (XMCD) analysis establishes that the cation ($Fe^{3+}$/$Co^{2+}$) redistribution occurs on tetrahedral and octahedral site in the electrically poled $CoFe_2O_4$/PMN-PT composite, confirming the coupling between magnetic and electric ordering in the composite. The magneto-electric coupling coefficient α vs dc magnetic field curves revealed hystereticbehavior and enhanced α values after electric poling, which originates from the strain induced modifications in the magnetic domains configuration of composite in the poled samples. These findings suggest that the existence of spin lattice coupling may leads to the mechanism of strong magneto-electric effects via strain interactions in $CoFe_2O_4$/PMN-PT composite.

**Keywords:** Multiferroic composites, spin-lattice coupling, XMCD, Raman Spectrosocpy, magneto-electric coupling.

1. Introduction

Recently, multiferroic composites exhibiting strong magneto-electric(ME) effects at room temperature have attracted tremendous attention for developing fascinating device concepts for spintronics and memory storage applications. The ME composites permit modification in the electric polarization by a magnetic field (direct ME effect) or the magnetization by an applied electric field (converse ME effect)[1,2]. Ferroelectric materials with high piezoelectric coefficients and ferromagnetic materials with high magnetostriction are suitable candidates to achieve strong ME effects in multiferroic composites, such as $NiFe_2O_4/Pb(Mg_{1/3}Nb_{2/3})O_3–PbTiO_3$(NFO-PMNPT), $PbZrTiO_3$-$CoFe_2O_4$ (PZT-CFO), $PbTiO_3$-$CoFe_2O_4$ (PTO-CFO), $BaTiO_3$-$CoFe_2O_4$ (BTO-CFO) etc.[3–5].

Owing to the correlated crystal and magnetic structures, the ME coupling in multiferroic composites, which is quantified as ME voltage coefficient ($\alpha$), depends mainly on the strain transfer at the interface between the ferroelectric/piezoelectric and ferromagnetic/magnetostrictive phase [5–8]. Generally, X-ray diffraction (XRD) and Raman spectroscopy are used to probe the induced strain lattice distortion and correlations between lattice and spin dynamics [9–10], as shown in our previous studies also on $NiFe_2O_4$/PMN-PT composites [11-12] wherein we observed modification in the crystal structure across the ferroelectric transition temperature of PMN-PT (owing to the strain transfer from PMN-PT to $NiFe_2O_4$). Further, the magnetic structure can also be modified due to the interfacial strain transfer from the ferroelectric phase to the magnetic phase [13]. Such strain induced mofications in crysal and magnetic structure in multiferroic composites help to to comprehend the mechanism of ME coupling.

In this work, we have investigated the spin-lattice correlations in strain coupled magneto-electric multiferroic $CoFe_2O_4/0.65Pb(Mg_{1/3}Nb_{2/3})O_3–0.35PbTiO_3$(PMN-PT) composite. PMN-PT is a robust ferroelectric material and exhibits structural phase transition (*P4mm* to $Fm\bar{3}m$ space group) at its Curie temperature ($T_c$)~ 450 K[14]. $CoFe_2O_4$ (CFO) is a known ferrimagnetic material with high magnetostriction ($\lambda$=200-400ppm) and becomes paramagnetic beyond~870 K [15]. The chosen materials have good interface linking and hence displays an effective strain transfer to the CFO phase upon the structural transition of PMN-PT.

We would like to emphasise that although the $NiFe_2O_4$ and $CoFe_2O_4$ exhibit nearly similar crystal structure (inverse spinel), the magnetic properties are quite different; $NiFe_2O_4$ is a soft ferrimagnets whereas $CoFe_2O_4$ is hard ferrimagnet material and its magneto-striction is higher as compared to $NiFe_2O_4$ [16-17]. It should be noted that magneto-striction is an important factor in determining the magneto-electric properties of strain coupled multiferroic composites.

Therefore it is appealing to probe strain induced changes in crystal and magnetic structure of the $CoFe_2O_4$ and the correlations between lattice and spin dynamics, which have not been studied so far in $CoFe_2O_4$ /PMN-PT composite samples.

## 2. Experimental

### 2.1 Synthesis

The nanocomposite of the molar ratio of 0.70(PMN-PT)-0.30(CFO) was prepared by mixing parent nanopowders of CFO and PMN-PT. All precursor chemicals were of analytical grade and used as received without further purifications. CFO nanopowder was prepared using a citric acid based sol–gel combustion method[18] and PMN-PT nanopowder was prepared using awet chemical method,as explained in detail elsewhere[19].The individually prepared powders of CFO and PMN-PT were mixed and grounded for 5 hrs. The fine grounded powder of the mixture was pelletized and sintered at 750° C for 10 hrs.

### 2.2 Used experimental technique

The crystal structures of the samples were analyzed bythe X-ray diffraction (XRD) technique using synchrotron source (E=15.431 keV). Raman spectroscopy measurements were carried out using Horiba JobinVyon, micro-Raman spectrometer (488 nm excitation laser), equipped with an edge filter for Rayleigh line rejection, and a CCD detector. The spectral resolution was ~1 $cm^{-1}$. The magnetic measurements were performed using Qunatum Design SQUID-VSM. The dielectric measurements were performed using a Novacontrol impedance analyzer. X ray absorption spectra (XAS) and X-ray magnetic circular dichroism (XMCD) were measured at the Fe and Ni $L_{2,3}$ edges at room temperature in total electron yield mode at polarized light soft x-ray absorption beamline BL1 at Indus-2 synchrotron source, RRCAT, Indore (INDIA). XMCD spectra were obtained at applied field of ±1T and fixed helicity of the incident photon beam. The magneto-electriccoupling coefficient measurements were performed by changing the bias magnetic field (1T) under application of $H_{ac}$= 10 Oe (at frequency = 10 kHz) produced using Helmholtz coils. The voltage output generated from the composite was measured by the lock-in amplifier (Stanford, SR532); the reference signal was taken from the signal generator feeding the Helmholtz coils.

### 3. Results and Discussion

#### 3.1 Crystallographic Phase analysis

Figure 1(a) showsRietveld refined XRD pattern of PMN-PT/CFO composite measures at 300 K. The refinement analysis reveals tetragonal structure (space group - P4mm) of PMN-PT and cubic inverse spinel ( space group - Fd-3m) of NFO in the composite.Room temperature crystal symmetry of the two phases in the composite is further confirmed by Raman spectroscopy, as shown in figure 1(b). The phononmodes $E_g+3T_{2g}+A_{1g}$ correspond to CFO inverse spinel structure with space group $O^7h$ ($Fd\bar{3}m$)[20]. The peaks observed for PMN-PT are mainly attributed to longitudinal optical (LO) modes i.e $A_1$ and E modes of tetragonal PMN-PT (Fig. 1(b)).[21]. The Raman modes of the composite are slightly shifted in the position as compared to the individual PMN-PT and CFO. The $E(LO_2)$ modes of PMN-PT phase shows a redshift, while $A_{1g}$ phonon mode of the CFO phase exhibits a blue shift. It indicates that CFO undergoes compression while PMN-PT lattice experiences tensile strain in the composite, concomitant with the XRD results where CFO peaks are shifted towards lower 2θ side and PMNPT reflections are shifted towards higer 2θ values, as shown in inset of Fig. 1(b). It divulges that PMN-PT and CFO are coupled via elastic interactions in the compositevia straintransfer through the interfaces between piezoelectric PMN-PT and magnetostrictive CFO.

#### 3.2 Temperature dependemnt XRD and Raman spectroscopy analysis

Across $T_C$, the structural phase transition of PMN-PTwill influence the adjacentlattice of CFO, which can be realized from the temperature dependent XRD measurements. Figure 2 (a) shows XRD patterns of the CFO/PMN-PT nanocomposite were measured at different temperatures 300 K, 400 K and 500 K (i.e across $T_c$ of PMN-PT). Figures2(b) and 2(c)show an enlarged view in the selected 2θ range of 16-19.5degree and 21.8-23.8 degree. Figure 2(c) indicates diffraction peaks corresponding to CFO (400) reflection and PMN-PT (200) reflection. Owing to the structural transition of PMN-PT at 450 K, it is observed that the doublet in (200) reflection of PMN-PT at 300 K corresponding to the tetragonal phase changes to a single Bragg peak P(200) at 500 K corresponding to cubic $Pm\bar{3}m$ phase, [22,23]. From the XRD analysis of CFO phase in composite (Figure 2 (b)), it is observed that the intensity ratio $I_{(220)}/I_{(222)}$= 4.25 at 400K changes considerably to 5.16 at 500 K. It should be noted that in inverse spinel ferrites e.g $NiFe_2O_4$, $CoFe2O4$, $I_{(220)}/I_{(222)}$ ratio depends on the cations occupying the A sites and B sites[24-27]. This observation reveals the modification in Co/Fe distribution at the T and O sites in CFO phase of the composite across $T_c$ of PMN-PT, which is further confirmed from the temperature dependent Raman spectra studies (Fig. 3 (a)). It is well known that $A_{1g}$ modes and $T_{2g}$ (2) mode are highly sensitive to the cations($Co^{2+}$/$Fe^{3+}$ ions) distribution at T and O-sites, respectively[24-27]. The mode position and line width of $A_{1g}$ and $T_{2g}(2)$ modes demonstrate deviation from usual anharmonic behaviour [28] and exhibit strong anomalies around $T_C \sim 450$ K of PMN-PT (Fig. 3(b)-(e)). These observations divulge lattice instabilities in CFO across $T_C$, which arise due to strain transfer through interface induced by the structural transformation in PMN-PT.

Any kind of change in the cation inversion in CFO affects the relative intensity of($A_{1g}(1)$+$A_{1g}(2)$) modes in (doublet at ~ 700 $cm^{-1}$)[24–26]. The relative intensity of $A_{1g}(1)$ and $A_{1g}(2)$ modes represents cation inversion, which can be written as $X_{Fe/Co}$, where X represents cation inversion factor. The parameter $X_{T/O}$ equals to 0 when there is no cation inversion and equals to 1 if the cation inversion is complete. Figure 4 (a) shows a plot of the relative intensity i.e $\frac{I_{A1g(1)}}{I_{A1g(1)+A1g(2)}}$(=$X_{T/O}$) as a function of temperature, which reveals $X_{T/O}$ to be nearly constant upto $T_C \sim 450$ K and it drops suddenly beyond $T_C$. This divulges that Co/Fe distribution in CFO phase of the composite is modified beyond 450 K, concomitant with the XRD studies (Fig. 2(b)). Figures 4 (b)&(c) present spectral deconvolution of the composite across the selected temperature region (300 K and 500 K) where the strongest anomalies are observed in Raman peak positions and line width. These modifications in CFO phase arise owing to the structural transition of PMN-PT at ~ 450 K, which causes a strain on the adjacent lattice of CFO in the composite and modifies its lattice. Earlier such kind of modification in cation distribution at the T and O sites was reported in pressure dependent studies on CFO [20,26,27].

*3.3 Correlation between Magnetiztion and phonon behaviour*

The temperature dependent magnetization was measured in field-cooled warming (FCW) mode under the applied magnetic field of 100 Oe, for CFO and CFO/PMN-PT composite, as shown in Fig. 5(a). CFO exhibits a characteristicferrimagnetic behavior with gradual increase in magnetization as the temperature is lowered below its magnetic ordering temperature 750 K [22].However, the CFO/PMN-PT composite shows an unusual drop in themagnetic moment around $T_c$-450 K of PMN-PT.The $T_c$ of PMN-PT is indicated by a peak in the temperature dependent dielectric constant around 450 K (Fig. 5(b)).The abrupt change in magnetization can be attributed to the lattice instabilities in CFO near $T_C$ of PMN-PT, which are quite evident from the phonon behaviour (frequencies and line width of $A_{1g}$ and $T_{2g}(2)$ phonon modes) near 450 K.

The correlation of anomalous phonon behaviour with magnetization can be better exemplified via spin-phonon coupling, caused by the phonon modulation of spin exchange integral[28,29]. Granado's et al. proposed that the phonon renormalization at magnetic ordering temperature is proportional to the nearest neighbour spin-spin correlation function and scales with the square of the sublattice magnetization.[28]Accordingly, the spin-phonon coupling can be evaluated by the spin correlation functions <*S*i ·*S*j> (where *S*i and *S*j are the localized spins at the *i*th and *j*th sites), which leads to

$$\Delta\omega(T) = \omega_{anh}(T) - \omega(T) \propto \left(M^2(T)/M^2_{max}\right) \qquad \ldots\ldots\ldots (1)$$

where M(T) is the average magnetization per magnetic ion, $M_{max}$ is the saturation magnetization. Figure 5(c)clearly shows that the temperature dependent phonon frequency shift $(\Delta\omega(T))$ of $T_{2g}(2)$ mode of CFO phase in the composite scaleswith the square of itsmagnetization $\left(M^2(T)/M^2_{max}\right)$, demonstrating the spin-phonon coupling in the composite unambiguously. The spin phonon coupling imprints its effct on the magneto-electric character also. TheME effect is evident from Figures 5 (a) and (b),where the concurrence between dielectric and magnetic ordering behaviour can be observed, indicating towards a possible ME coupling.

*3.4  Evidance of magnetoelectric coupling analysis*

In order to confirm the existence of ME coupling in CFO/PMN-PT composite, the magnetization as a function of magnetic field (M−H) was measured before and after electrical polingat an applied electric field of E= +40 KV/cm, as shown in Figure 6(a). The applied electric field was above the coercive field ($E_C$ ~ 23 kV/cm) of CFO/PMN-PT composite at room temperature (inset Fig. 5(b)). The conductivity and loss tangent are of the order $10^{-9}$ $ohm^{-1}cm^{-1}$ and 0.21 at frequency 1 kHz, respectievely (Fig. 6(b)). It is observed that after electric poling, saturation magnetization decreases from 44 to 39.5 (± 0.2) emu/g and coercivity of the composite increases drastically. Poled composite sample exhibits non-volatile changes in the magnetization i.e it does not revert back to its initial value after removal of electric field.

The electric field leads to the ferroelectric dipoles alignment along the electric field direction and consequently theferroelectric/piezoelectric PMN-PT experiences elongation in the field direction which produces stress in the magnetostrictive CFO phase in the composite sample. The induced stress modifies the crystalline arrangement as well as spin and orbital degrees of freedom [30, 31]. Hence the electric field induced straincan alter the magnetization of the composite. Also, the electric field causes accumulation of electrons/holes at the interface of PMN-PT and CFO, which would attract/repel the oxygen vacancies at the CFO/PMN-PT interfaces. This leads to redistribution of oxygen vacancy in the poled samples, as described in the schanmatic of the Fig. 6(c).The magnetic structure of CFO consist of two anti-ferromagnetically ordered sublattices i.e. A and B, corresponding to the tetrahedral ($T_d$) and octahedral ($O_h$) sites, respectively with the $Co^{2+}$ cations occupying $O_h$ sites and the $Fe^{3+}$ cations occupying equally both the $T_d$ and $O_h$ sites [32,33]. The redistributions of oxygen vacancy can modify the Ms and Hc values via changes in the superexchange interactions of A-O-B, A-O-A and B-O-B and redistribution of local valences of cations ($Fe^{2+\backslash 3+}$, $Co^{2+\backslash 3+}$) at O and T sites in the CFO [31-33]. The redistiution of the cations ($Fe^{3+}$, $Co^{2+}$) on the tetrahedral ($T_d$) and octahedral ($O_h$) sites is shown schematically in Fig. 7. Thus the altered magnetization in the poled CFO/PMN-PT may arise due to the collective effect of electric field induced strain of PMN-PT, and charge modulation at the CFO/PMN-PT interfaces upon electric poling.

In this context, XMCD technique is used to study the spin structure (cationic sites distribution vis a vis magnetic moment at O and T sites) in poled and unpoled composite

CFO/PMN-PT samples. XMCD provides detailed fingerprint of spinel structure where each specific cation (Fe/Co) generates a unique XMCD signature that is determined by its valence state (number of d electrons), site symmetry (i.e., $T_d$ or $O_h$), and magnetization (local moment) direction[34]. Figures 8(a-b) and (c-d) present XAS spectra of Fe and Co $L_{2,3}$ edges with the helicity of incident X ray parallel ($\sigma^+$) and antiparallel ($\sigma^-$) to the direction of magnetization and the corresponding XMCD signals for unpoled and poled composite samples, respectively. The $L_3$ edge Fe XMCD signal (Fig. 8(a)) displays mainly three peaks where the first very small negative peak (A1) indicates contribution of $Fe^{+2}$ at $O_h$ site, next peak (A2) arises from $Fe^{+3}$ at $T_d$ site and third negative peak (A3) is attributed to the $Fe^{+3}$ ions at $O_h$ site[35].The oppiste sign of A1 and A3 with respect to A2 implies that the magnetic moments of the $Fe^{3+}$ ions in $T_d$ symmetry are coupledantiferromagnetically with the $Fe^{3+}$ ions in $O_h$ symmetry sites. A C2 feature with negative XMCD signal of L$3$ edge of Co edge (Fig. 8(b)) suggests that $Co^{2+}$ ions occupy $O_h$ sites. Appearance of feature C1 suggests that a small fraction of Co ions are residing at the $T_d$ site also.The observed characteristic features of XMCD signal for the unpoled composite match well with CFO [35].

A linear combination of the simulated XMCD spectra using charge transfer multiplet calculations (CTM4XAS code[36]), is used to quantify the $Fe^{2+/3+}$ cations distribution at tetrahedral ($T_d$) and octahedral ($O_h$) sites. The analysis reveals 72% $Fe^{3+}$ ions at $O_h$ sites, 20% $Fe^{3+}$ions at $T_d$ sites and a small fraction (8%) of $Fe^{+2}$ions occupy $O_h$ sites, for the unpoled CFO/PMNPT composite. While for the poled composite samples, 59% $Fe^{3+}$ ions are at $O_h$ sites, 36% $Fe^{3+}$ ions at $T_d$ sites and a small fraction (5%) of $Fe^{+2}$ ions occupy $O_h$ sites. It suggests that a fraction of $Fe^{3+}$ ions is transferred from $T_d$ to $O_h$ site after electric field poling and a fraction of $Fe^{2+}$ ions gets reduced at the $O_h$ site after poling. The comparison of the Co $L_{2,3}$ edges XMCD signal for poled and unpoled PMN-PT/CFO reveals that the C2 peak intensity is reduced and a small positive peak (C1) at 778 eV in unpoled sample is missing in the poled sample. These observations suggest that while CFO in unpoled sample possesses mixed inverse spinel structure, it gradually moves towards inverse spinel structure upon electric poling, which also modifies the magnetic properties of CFO in poled CFO/PMN-PT samples. This further suggests that the electric field poling leads to the redistribution or introduction of oxygen vacancies in the CFO through CFO/PMN-PT interfaces.

Further, the orbital (ml) and spin (ms) contributions to the total magnetic moment per cation for the poled and unpoled composite samples can be estimated by sum rule[37,38], which allow the evaluation of average total magnetic moment (m = ml + ms) of the Co and Fe ions.

$$m_l = -\frac{4}{3} N^h \frac{q}{r} \quad \ldots \ldots \ldots \ldots \ldots \ldots (2)$$

$$m_s + <T_z> \geq -N^h \left(\frac{6p-4q}{r}\right) \ldots \ldots \ldots \ldots \ldots (3)$$

where $m_l = -\frac{4}{3} N^h \frac{q}{r}$

$$r = \int_{L_{2,3}} (\sigma^+ + \sigma^-) dE, p = \int_{L_3} (\sigma^+ - \sigma^-) dE, q = \int_{L_{2,3}} (\sigma^+ - \sigma^-) dE$$

$N^h$ represents the number of 3d band holes, p is the integral of the XMCD signal for the L3 edge, q is the integral of the XMCD signal for the L3 and L2 edges, and r is the integral of the average XAS spectrum. The equation no. (2) is directly used to calculate the $m_l$, while an effective $m_s$ is estimated (using equation no. (3)), by including the expectation value of the magnetic dipole operator ⟨$T_z$⟩, which is negligible for transition metal compounds at room temperature and hence neglected in the present case[39].

In unpoled sample, the value of Co $m_S$ is estimated to be 0.32$\mu_B$ and $m_L$=0.03 $\mu_B$, yielding the magnetic moment contribution due to Co ion $M_{Co}$=0.35$\mu_B$. Similarly for the Fe ions, $m_S$= 3.1 $\mu_B$ and $m_L$=0.19 $\mu_B$, and total $M_{Fe}$ =3.29 $\mu_B$. Hence, the net magnetic moment for the unpoled sample is estimated to be ~ 3.64 $\mu_B$. The total moment is slightly higher than expected values for CFO (3 $\mu_B$) having inverse spinel structure, which is consistent with the somewhat mixed spinel structure of CFO as evident from the XAS data analysis. For the poled samples, we estimate $M_{Co}$= 2.46 $\mu_B$(($m_S$=2.10 $\mu_B$, $m_L$=0.36$\mu_B$ $\mu_B$)), and $M_{Fe}$=0.73 $\mu_B$(($m_S$= 0.70 $\mu_B$, $m_L$=0.03 $\mu_B$)), yielding total moment 3.19 $\mu_B$. This is consistent with our XAS observation that upon electric poling, the system has a tendency to adopt an inverse spinel structure, wherein the magnetic moment of CFO is largely attributed to the Co ions as Fe ions moments cancel each other. It is noted that after poling the relative contribution of spin and orbital magnetic moment changes considerably. These results establish coupling between electric and magnetic ordering in the CFO/PMN-PT composite samples, as the application of electric field causes redistribution of cations at $O_h$ and $T_d$ sites.

Moreover, the coupling can be quantified in terms of the magnetoelectric coefficient (α) measured as a function of in-plane $H_{dc}$ at an off-resonance frequency of 10 kHz under superimposed in-plane ac magnetic field ($H_{ac}$) of 10 Oe, as shown in Figures 9 (a) and 9 (b).

Interestingly, the unpoled composite also reveals ME coefficient (Fig. 9(a)) as the two phases are strain coupled with each other. The poled composites show enhanced α value, clear hysteresis behaviour with coercive field of ~ 4000 Oe as well as remnant α value of 40 mV/cm$^{-1}$Oe$^{-1}$ (Fig. 9(b)), establishing self-biased ME effect [40] in CFO/PMN-PT nanocomposite.The self-biased hysteretic behavior can be explained from effective magnetostriction (λ) behaviour[41] (Figs 9(c) and (d)), which is estimated by integrating α with respect to the H$_{dc}$ (as α is propotional to piezomagnetic coefficient (q=dλ/dH$_{dc}$)). For the unpoled sample, λ remains unchanged at H$_{dc}$=0 under small variation of magnetic field (δH$_{ac}$) (Fig. 9(c)) suggesting that magnetic domain cannot move easily in this condition and restrict domain wall migration. Whereas,poled composite exhibits asymmetric λ curves where λ has a finite value (with δH$_{ac}$) at H$_{dc}$=0 (shown by dotted lines in Fig. 9(d)), implying easy domain rotation under δH$_{ac}$ or δΓ$_{ac}$ (Γ$_{ac}$ = stress)and henceeffectivestrain generation/transfer[43]. Indeed, the interactions between built-in bias ((H$_{int}$) and H$_{dc}$cause shift and asymmetrical behaiour of effective λ, resulting in self-biased ME effects[42].The effective dc magnetic bias (H$_{Eff}$=H$_{int}$ + H$_{dc}$) in CFO/PMN-PT composite strongly depends upon the relative orientation of H$_{int}$ and H$_{dc}$. This causes different maximum α value at different H$_{Eff}$ during the reverse and forward sweep[41,42]. Hence the ME hysteretic behaviour w.r.t external H$_{dc}$ is governed bythe piezomagnetic coefficient's dependence on the H$_{Eff}$.

From these observations it transpires that whether the strain is induced via ferroelectric transitions or applied electric field, it is effectively transferred to the adjacent magnetic phase through interfaces and modify its crystal and magnetic structure, resulting in the strong coupling beween electric and magnetic order parameters in such composites.

**Conclusion**

In conclusion, we have investigated the strain induced magneto-electric effects and the modification in the structure of individual CoFe$_2$O$_4$ and PMN-PT due to strain transfer across the interfaces. It is revealed that the interfacial strain in CFO phase developed across the ferroelectric/structural transition (Tc ~ 450 K) of PMN-PT causes local lattice instabilities in CFO lattice and induces spin-phonon coupling. The room temeprarture magneto-electric coupling was evidenced by measuring M-H loops for the unpoled and electrically poled CFO/PMN-PTcomposite. XMCD studies reveal modified spin structure in the electric poled composite at room temeperature. Interestingly, electrically poled composites reveal improved

ME coefficient α vlaues with self biasing ME effects as compared to un-poled composites, owing to the modification in the magnetic domain structure of composite after electric poling.

**Acknowledgement:** The authors would like to acknowledge (CSIR), New Delhi, India for financial support. The authors thank Mr. Manvendra Narayan Singh for assistance in high temperature XRD measurement, Mr. Rakesh Shah and Mr. Avnish Wadikar for XMCD measurement at Indus beamline, RRCAT, Indore.

**Conflict of interest:** The authors have no conflicts to disclose.

**Figure captions:**

**Figure 1.** (a)Room temperature rietveld refined XRD pattern of CFO/PMN-PT composite at, (b) room temperature Raman spectra of pure CFO, pure PMN-PT and composite CFO/PMN-PT, inset shows enlarged view of XRD pattern of pure CFO, pure PMN-PT and CFO/PMN-PT composites in a selected 2θ range of 30.7–36.7 degree.

**Figure 2**. Rietveld refined comparative XRD pattern of CFO/PMN-PT composite collected at different temperature i.e 300 K, 400 K and 500 K; (b)enlarged view in the selected 2θ range of 15.7–19.7 degree; (c) enlarged view in the selected 2θ range of 21–25 degree.

**Figure 3.** (a)Temperature dependent Raman spectra of CFO/PMN-PT composite; emperature dependence of the phonon mode (b-c) positions and (d-e) width of T-site (A1g) and O-site (T2g(2)) modes observed for composite CFO/PMN-PT.

**Figure 4.** Temperature evolution of the intensity ratio $\frac{I_{A1g(1)}}{I_{A1g(1)}+A1g(2)}$ of $A_{1g}(1)$ and $A_{1g}(2)$ phonon modes for CFO/PMN-PT composite in the temperature range 300 K-600 K; (b-c) temperature dependent spectral deconvolution for selected frequency range of CFO/PMN-PT composite at 300 K and 500 K. The points show experimental data and the solid line represents the results of fitting of the phonon modes by multiple Lorentzian line shape function.

**Figure 5.** Temperature dependent magnetization curve measured at H=100 Oe for pure CFO and CFO/PMN-PT composite; (b) temperature dependent dielectric constant measured at various frequencies for composite, inset shows the polarization vs electric field , inset show (PE) loop) for CFO/PMN-PT composite(c) comparison of$(M(T)/Mmax)^2$ and $(\Delta\omega(T))$ as a function of temperature.

**Figure 6.** (a) Room temperature magnetization vs magnetic field (M−H) loops measured for CFO/PMN-PT compositebefore and after electrical poling; (b) frequency depenent

condictivity and loss tangent curves fot the unpoled CFO/PMN-PT composite; (c) schematic diagram for electric poling of CFO/PMN-PT composite; (d) unpoled state i.e electric field (E=0); (d) poled state ubder application of E= 40kV/cm.

**Figure 7.** Schematic representation of cations ($Fe^{3+}/Co^{2+}$) restribution in normal spinel and mixed spinel structure of CFO.

**Figure 8.** XMCD signal measured at the Fe (a) and Co (b) $L_{2,3}$ edges, for unpoled composite samples; (c)XMCD signal for Fe and (d) Co$L_{2,3}$ edges, for electrically poled composite sample.

**Figure 9.** Magnetic DC field dependent magnetoelectric coefficient (α) for (a) unpoled CFO/PMN-PT, (b) poled CFO/PMN-PT, (c) Integral values of α with respect to the $H_{bias}$ for unpoled CFO/PMN-PT, (d) poled CFO/PMN-PT.

Figure 1

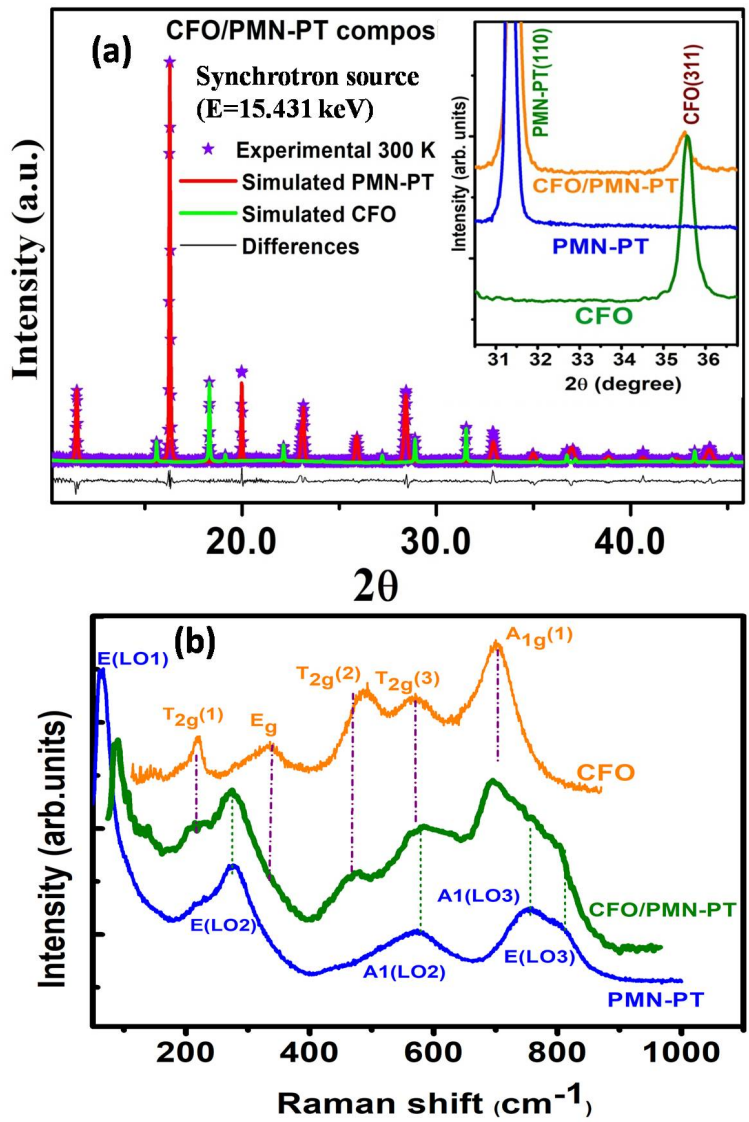

**Figure 2**

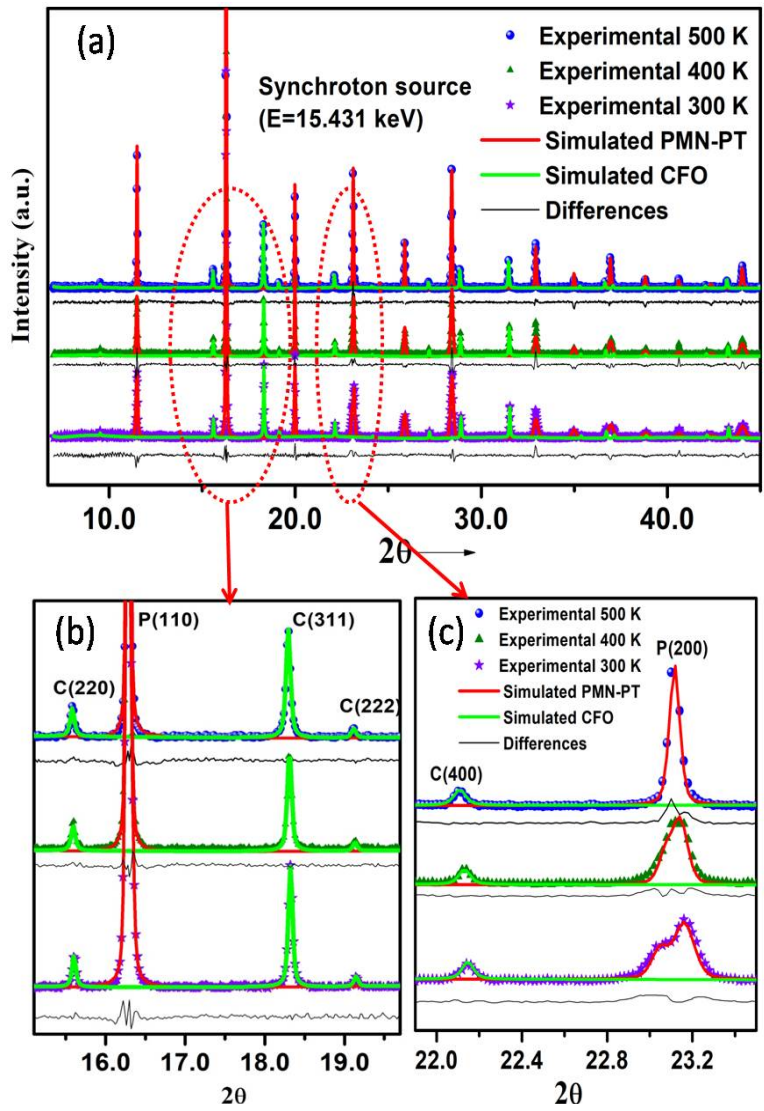

**Figure 3**

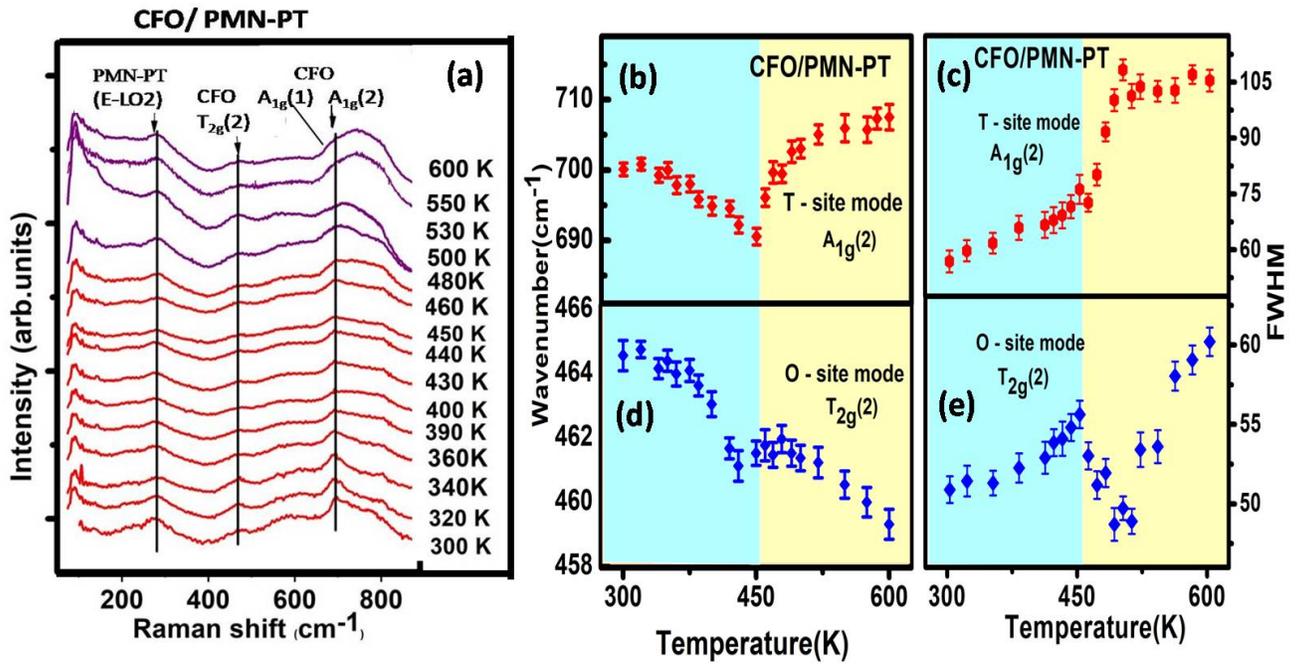

**Figure.4**

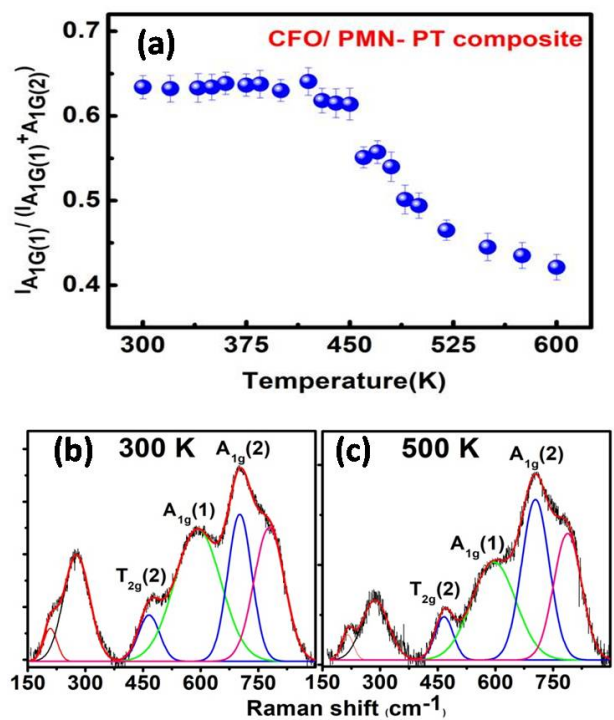

Figure.5

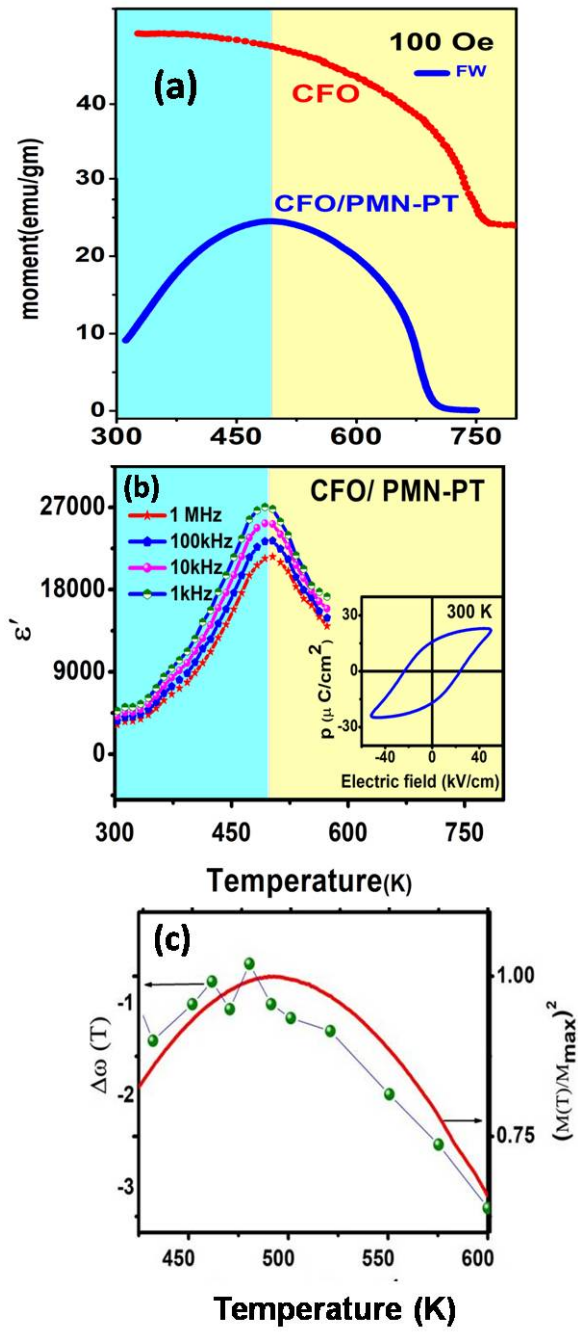

**Figure.6**

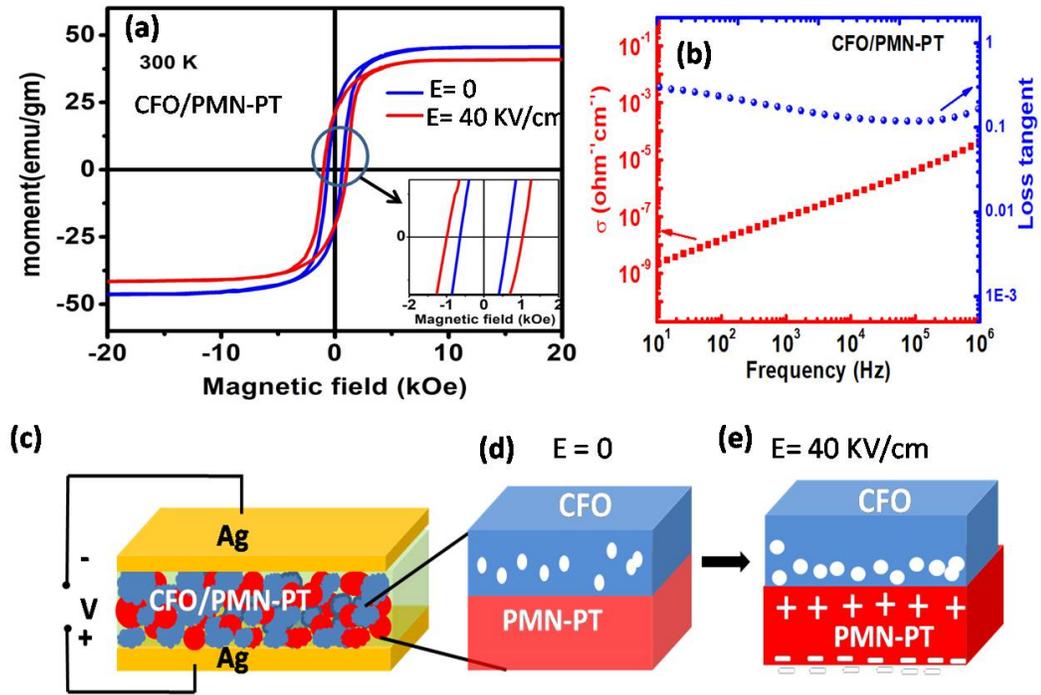

Figure 7

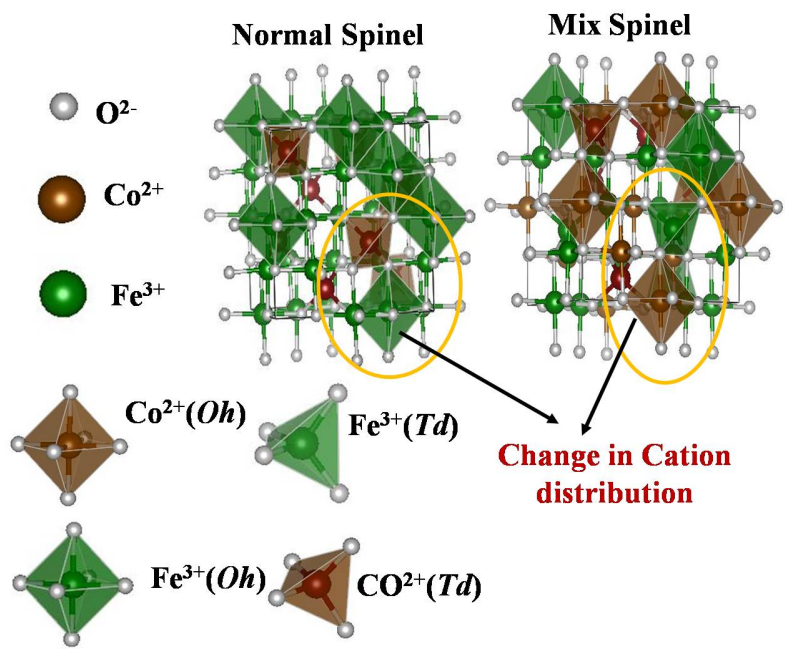

Figure 8

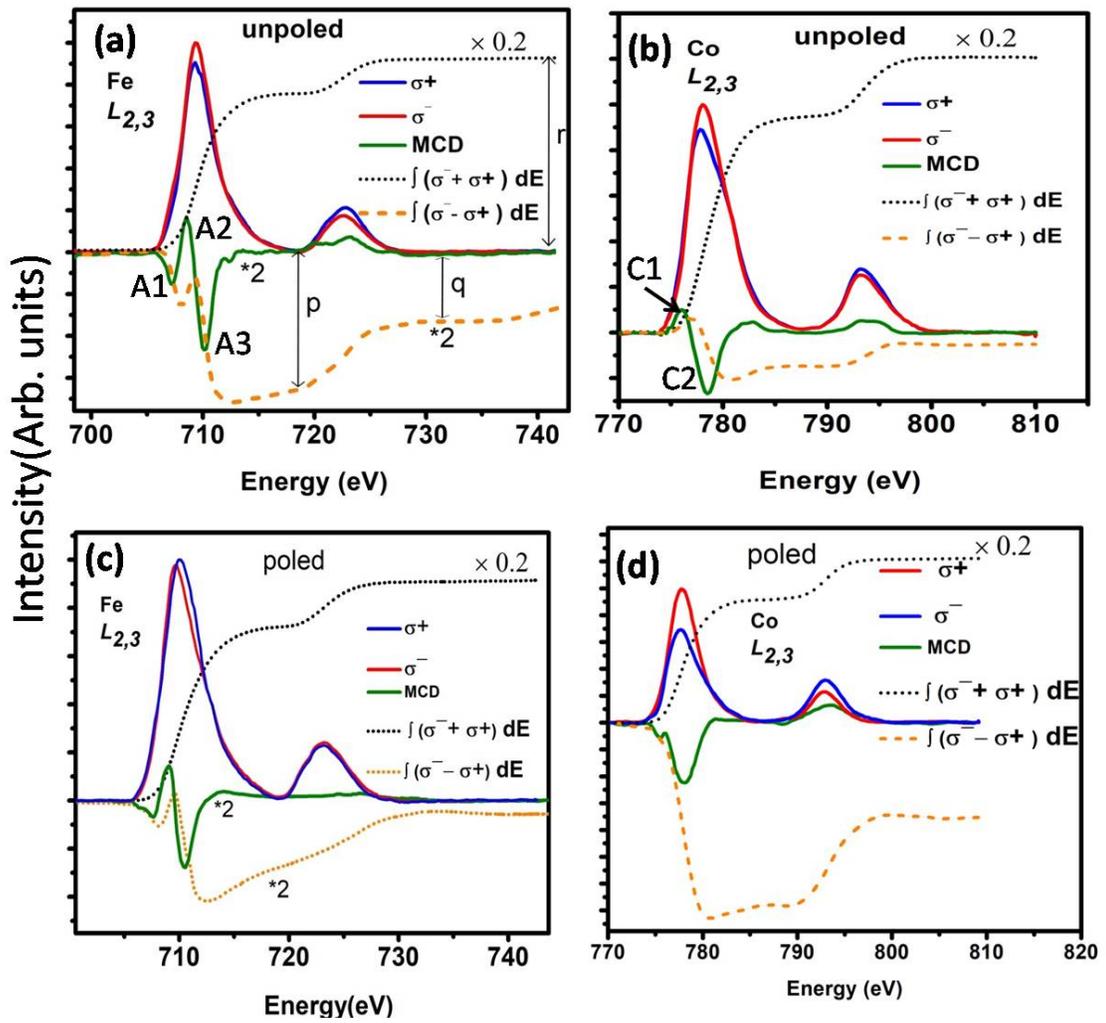

**Figure.9**

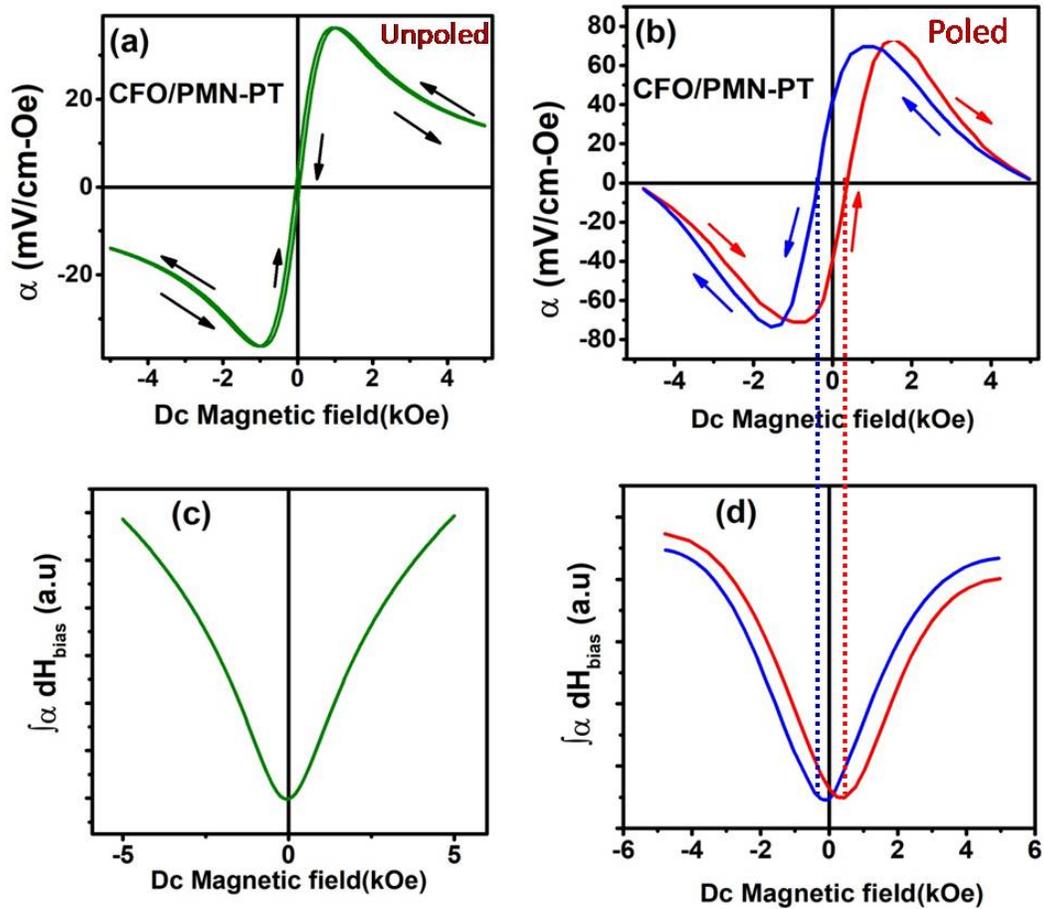